\documentclass{aip-cp}

\pdfoutput=1

\usepackage[numbers]{natbib}
\usepackage{rotating}
\usepackage{graphicx}

\usepackage{amsmath}

\newcommand{\ttbar}{\ensuremath{\mathrm{t}\overline{\mathrm{t}}}}
\newcommand{\mt}{\ensuremath{\mathrm{m}_{\mathrm{t}}}}
\newcommand{\mtbar}{\ensuremath{\mathrm{m}_{\overline{\mathrm{t}}}}}
\newcommand{\mw}{\ensuremath{\mathrm{m}_{\mathrm{W}}}}
\newcommand{\mtfit}{\ensuremath{\mathrm{m}_{\mathrm{t}}^{\text{fit}}}}
\newcommand{\mtreco}{\ensuremath{\mathrm{m}_{\mathrm{t}}^{\text{reco}}}}

\newcommand{\mwreco}{\ensuremath{\text{m}_{\mathrm{W}}^{\text{reco}}}}
\newcommand{\rbq}{\ensuremath{\mathrm{R}_{\mathrm{bq}}^{\text{reco}}}}
\newcommand{\mbl}{\ensuremath{\mathrm{m}_{\mathrm{l}\mathrm{b}}^{\text{reco}}}}
\newcommand{\pt}{\ensuremath{p_{\mathrm{T}}}}

\begin{document}

\title{Measurements of the Top Quark Mass at ATLAS and CMS} 



\author[cornell]{N. Mirman (for the ATLAS and CMS collaborations)}
\affil[cornell]{Cornell University, Ithaca, NY 14850, USA}%



\maketitle

\begin{abstract}
We present recent measurements of the top quark mass by the ATLAS and CMS experiments in the \ttbar\ lepton+jets, all-hadronic, and dilepton channels.  In addition, we present a measurement using a topology enriched in t-channel single top events.  The analyses include observables whose sensitivity to the top mass is calibrated using Monte Carlo simulation before they are utilized to extract the value of \mt\ in data.
The measurements outlined here enter into recent combinations by ATLAS and CMS that yield a sub-GeV precision on the top mass.
\end{abstract}

\section{INTRODUCTION}
\label{sec:introduction}
The top quark mass is a fundamental parameter of the Standard Model (SM), and its determination is an important part of the physics program at the Large Hadron Collider (LHC).  The top mass is an input into global electroweak fits that probe the self-consistency of the SM \cite{gfitter}, and its value has implications for the stability of the SM vacuum \cite{vacuumstability}.  Here, we give an overview of several recent top mass measurements by ATLAS \cite{atlas} and CMS \cite{cms} using proton-proton collisions from Run 1 of the LHC.  The analyses outlined here include observables whose sensitivity to the top mass is calibrated using Monte Carlo simulation before they are utilized to extract the value of \mt\ in data.
At the time of the LHCP 2015 conference, the most precise combination of measurements carried out at the LHC gave \mbox{$\mt = 172.38 \pm 0.66$ GeV} \cite{cms-comb}, achieving a precision of about 0.4\% on the top mass.

At the LHC, top quarks are predominantly produced in pairs via the strong interaction.  The $\mathrm{t}$ and $\overline{\mathrm{t}}$ then decay to $\mathrm{W}^+\mathrm{b}\mathrm{W}^{-}\overline{\mathrm{b}}$ almost 100\% of the time, yielding three possible final states -- \textit{all-hadronic}, \textit{lepton+jets}, and \textit{dilepton} -- determined by the subsequent W decays.  The all-hadronic channel has the largest branching ratio, but the selection of an all-jets final state lends itself to large QCD multijet backgrounds.  The lepton+jets channel typically provides the most precision for top mass measurements due to its balance between branching ratio and clean single-lepton selection.  The dilepton channel is the cleanest in term of physics backgrounds, but reconstructing the \ttbar\ system in this channel is complicated by the presence of two neutrinos in the final state.  In addition to \ttbar\ event topologies, the top quark mass has also been measured in events enriched with weakly-produced t-channel single tops.  Here, mass measurements probe a complementary regime in color flow, production energy scale, and final state event topology.

\section{CMS MEASUREMENTS}
\label{sec:CMS}
\begin{figure}
  \includegraphics[width=0.32\textwidth]{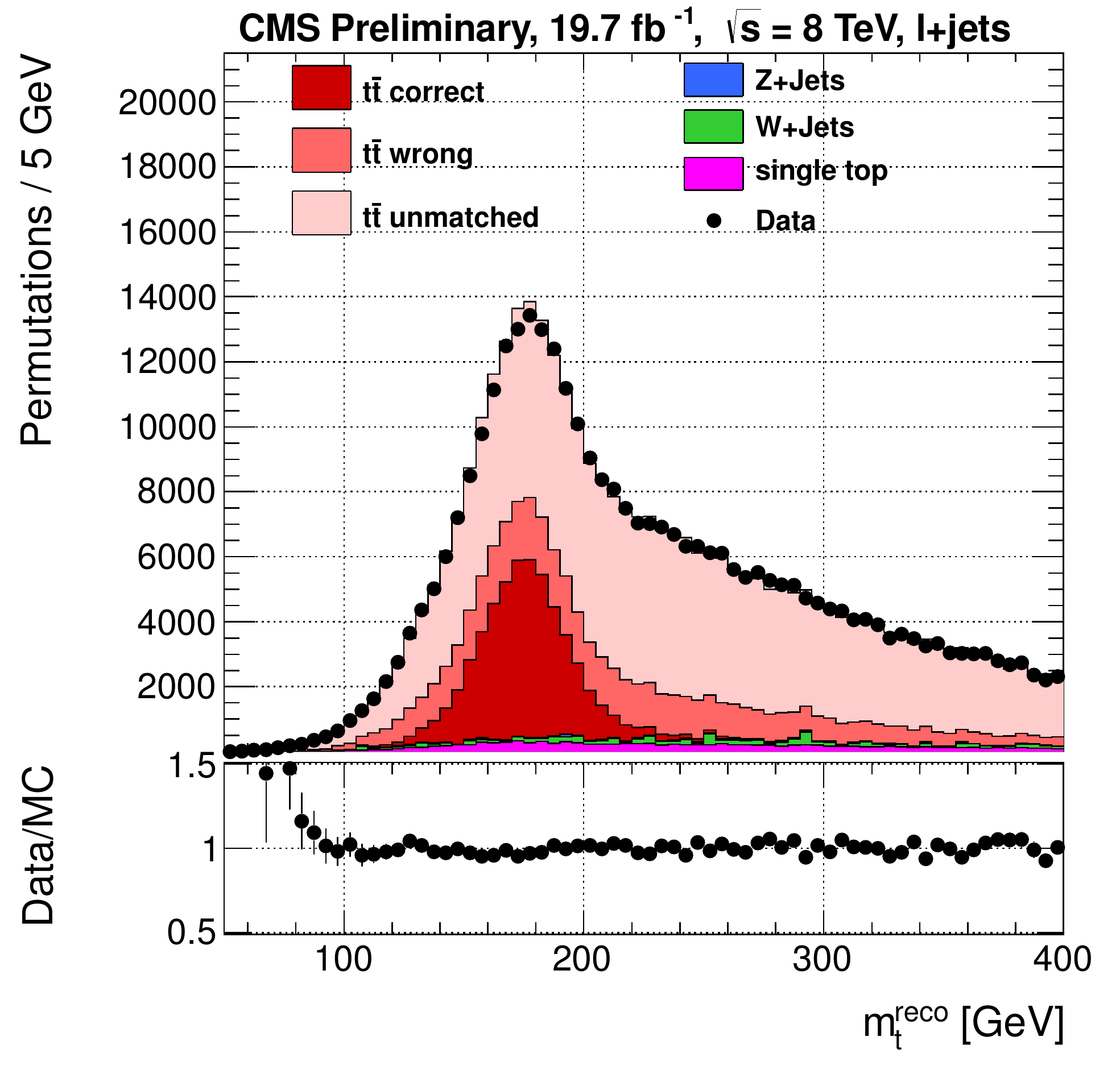}%
  \includegraphics[width=0.32\textwidth]{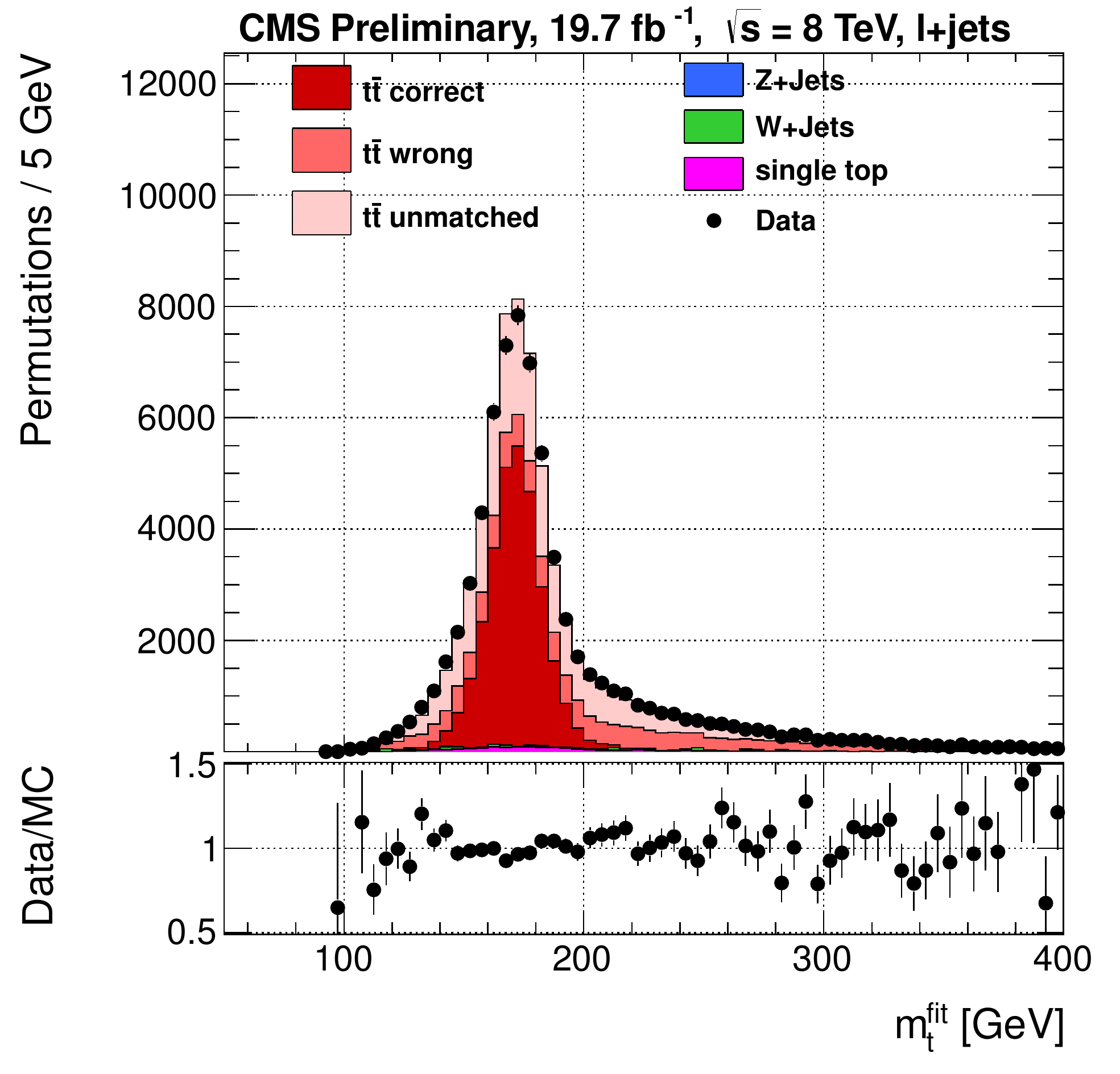}%
  \includegraphics[width=0.32\textwidth]{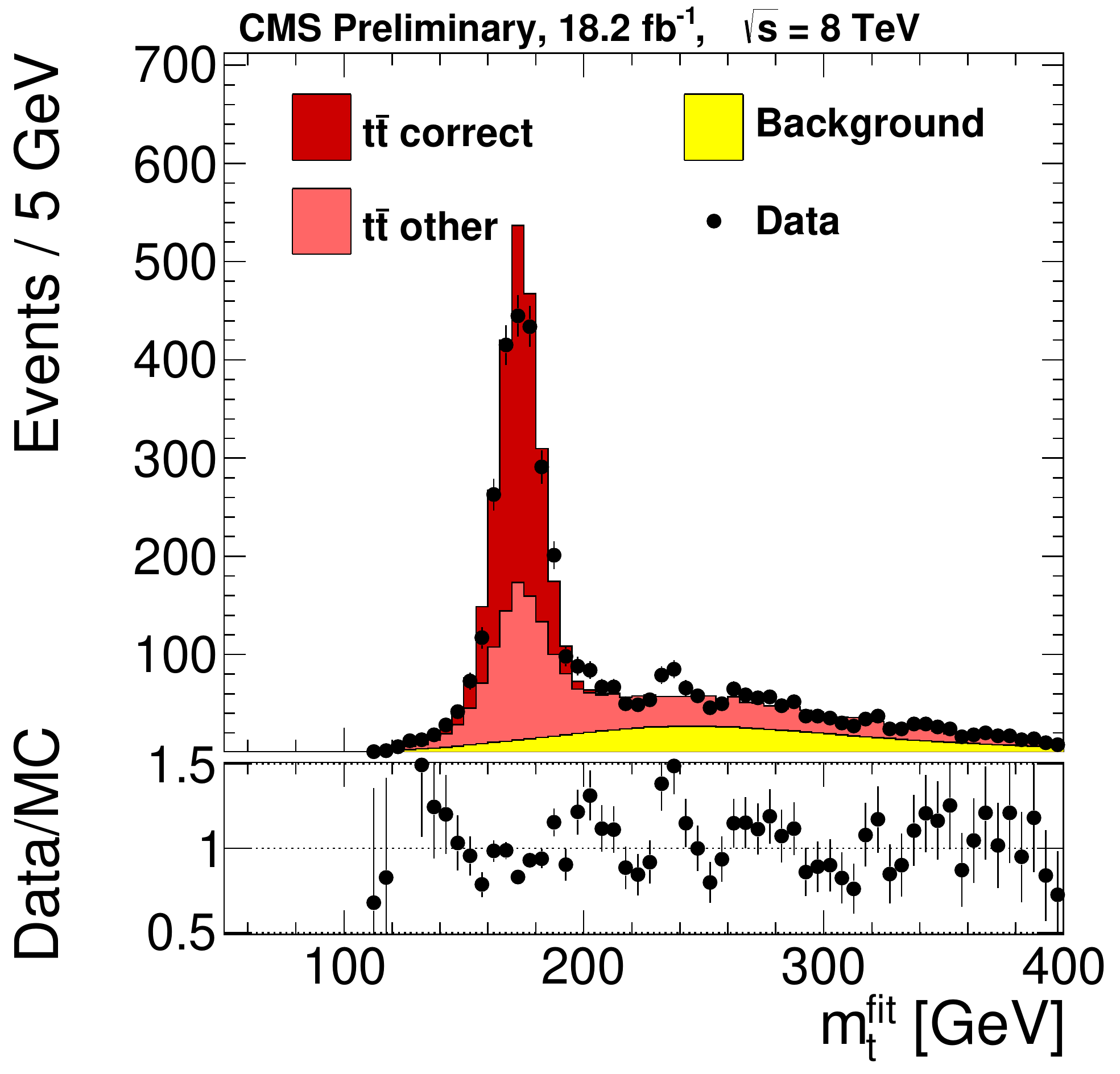}%
  \caption{Reconstructed top quark mass (left) before and (center) after the kinematic fit in the CMS lepton+jets channel.  As a result of the fit, the fraction of correct jet-parton permutations is increased from 13\% to 42\% and the resolution on the reconstructed \mt\ is improved.  (Right) reconstructed top mass in the all-hadronic channel, with the data-driven background estimate shown in yellow \protect{\cite{cms-ljets,cms-allhad}}.}%
  \label{fig:cms_kinfit}
\end{figure}

In the lepton+jets channel, events are selected from the full CMS 8 TeV dataset with an integrated luminosity of 19.7 fb$^{-1}$ \cite{cms-ljets}.  Events are required to contain a single isolated electron or muon, at least 4 jets, and exactly 2 b tags.  One challenge in the lepton+jets final state is the assignment of final state jets to their corresponding parent tops.  A kinematic fit is used to help mitigate these combinatoric effects in jet-parton assignment and to improve the resolution on the reconstructed top mass.  The general strategy of the kinematic fit is to vary the reconstructed objects within their respective resolutions, while imposing the mass constraints $\mw = 80.4$ GeV and $\mt = \mtbar$.  Then, a cut is placed on the $\chi^2$ probability to remove jet-parton assignments that are not likely to be correct.  The effect of this procedure is demonstrated in Figure~\ref{fig:cms_kinfit}.  Here, the improved \mtfit\ distribution is a result of the kinematic fit and $\chi^2$ probability requirement.

To reduce the effect of uncertainties due to the jet energy scale (JES) on the final result, jets reconstructed from the decay $\mathrm{W}\rightarrow \mathrm{jj}$ are used to perform an in-situ calibration of the JES.  Here, the JES enters into a 2D likelihood fit as a scale factor multiplying all jet four-vectors.  In the fit, the value of this scale factor is determined simultaneously with \mt.  Variations in the JES parameter affect both the \mtfit\ and \mwreco\ distributions, but its value is determined mostly by the latter, where the W mass is a powerful constraint on the dijet system.

\begin{figure}
  \includegraphics[width=0.32\textwidth]{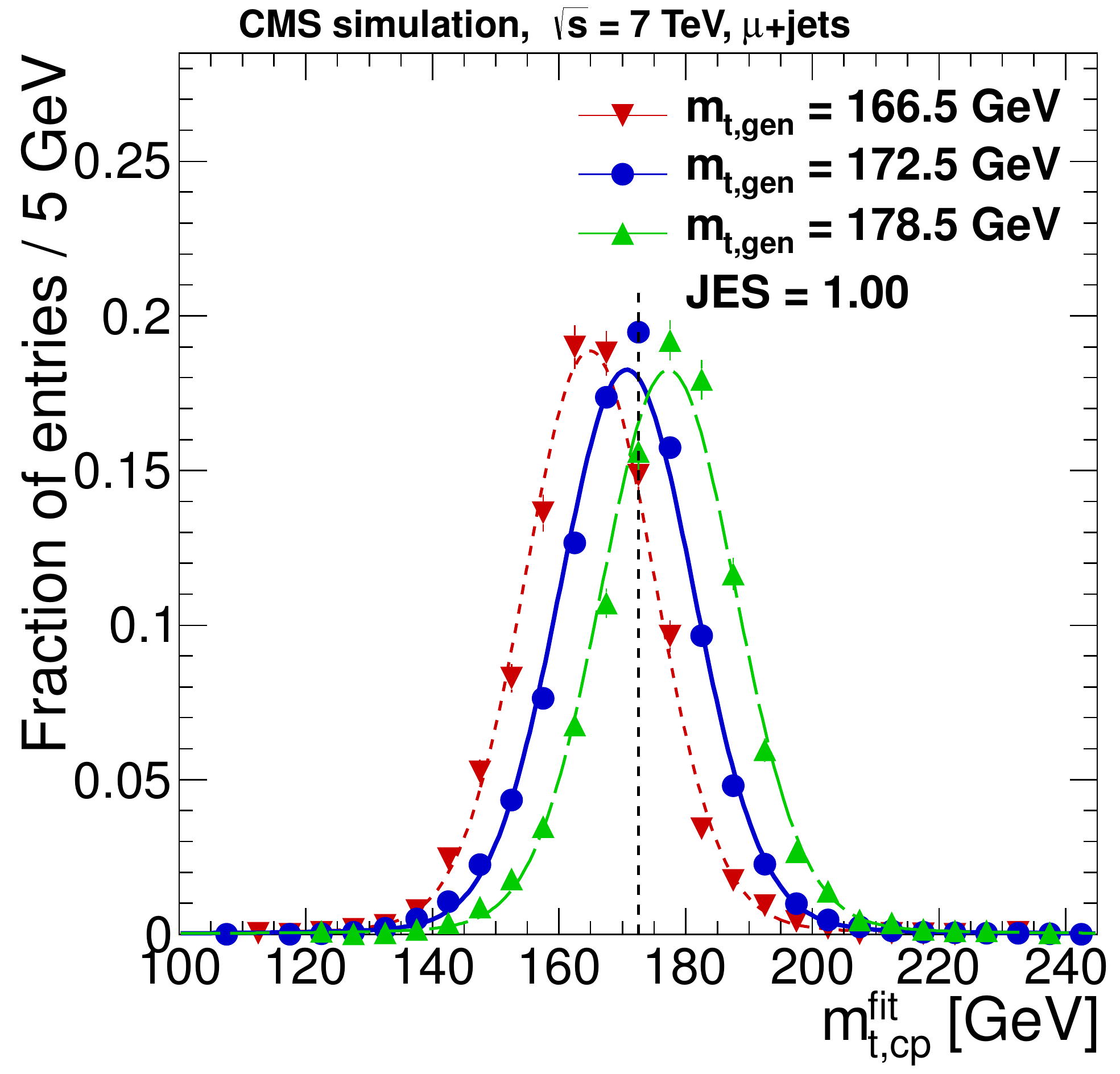}%
  \includegraphics[width=0.32\textwidth]{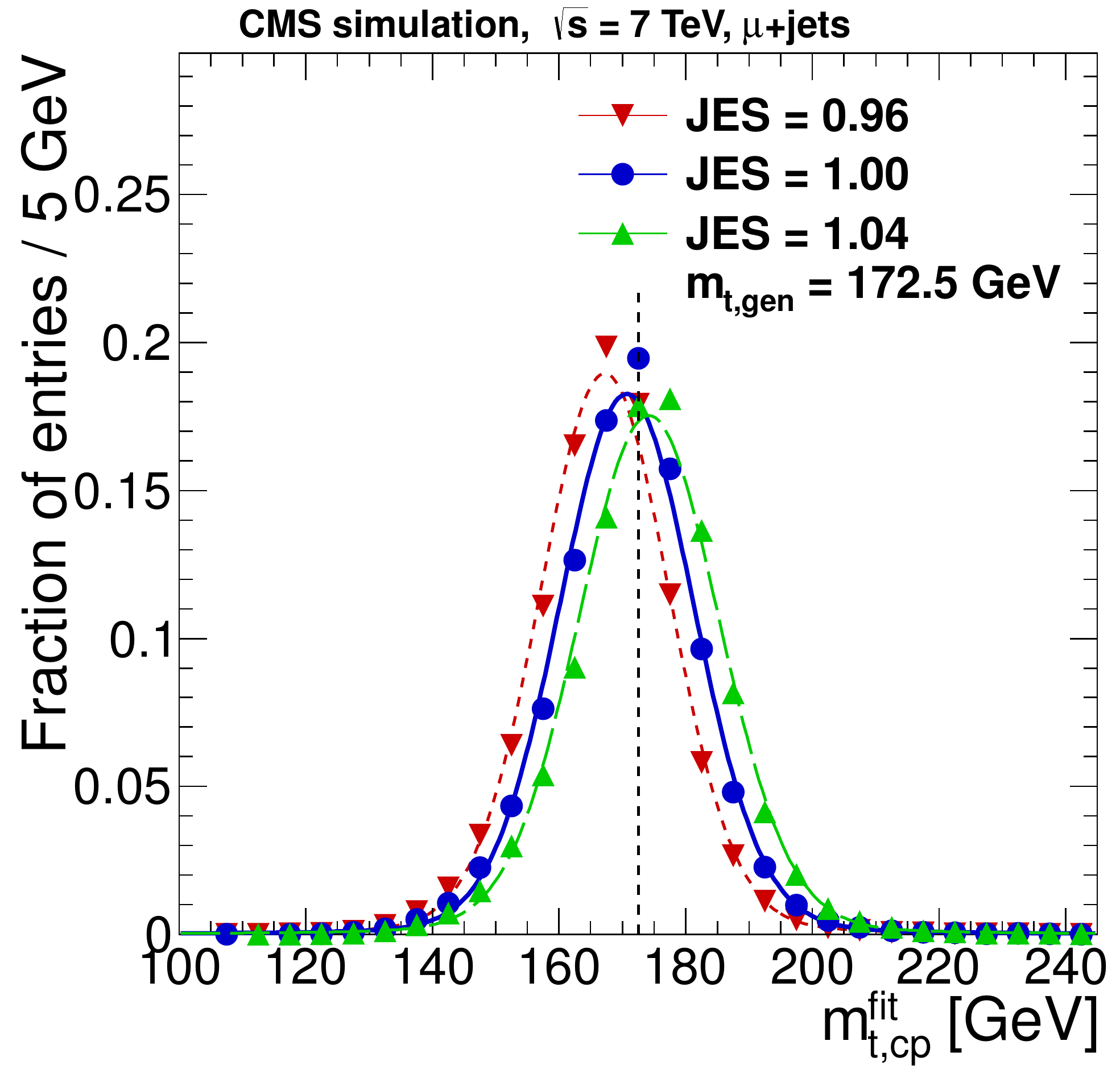}%
  \includegraphics[width=0.32\textwidth]{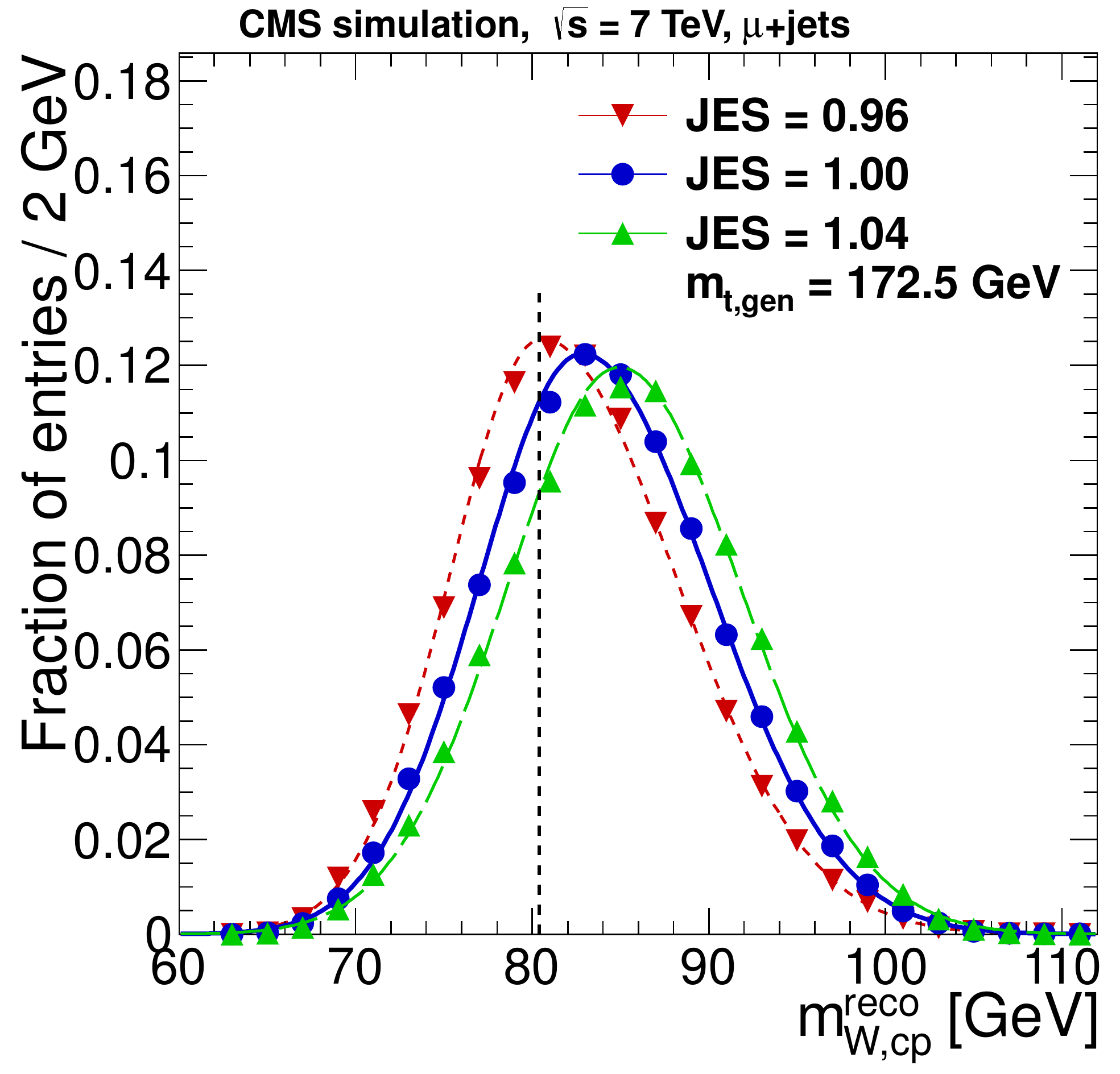}%
  \caption{Template shapes for the (left, center) \mtfit\ and (right) \mwreco\ distributions for correct jet-parton permutations in the CMS lepton+jets channel.  The \mtfit\ distribution is sensitive to both the \mt\ and JES fit parameters, while the \mwreco\ distribution is sensitive only to the JES \protect{\cite{cms_ljets_5fb}}.}
  \label{fig:cms_ljets_templates}
\end{figure}

\begin{figure}
  \includegraphics[width=0.32\textwidth]{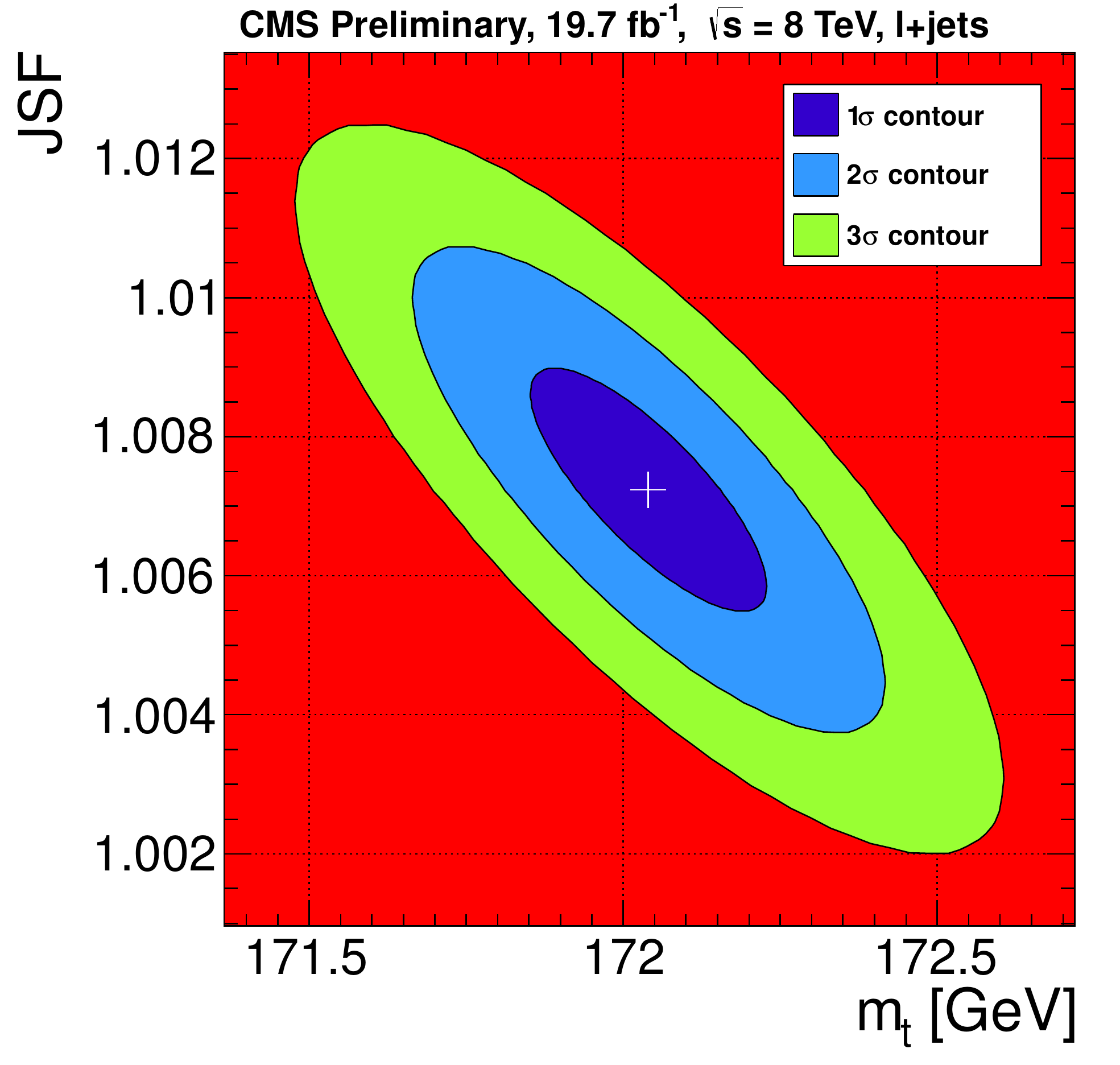}%
  \includegraphics[width=0.32\textwidth]{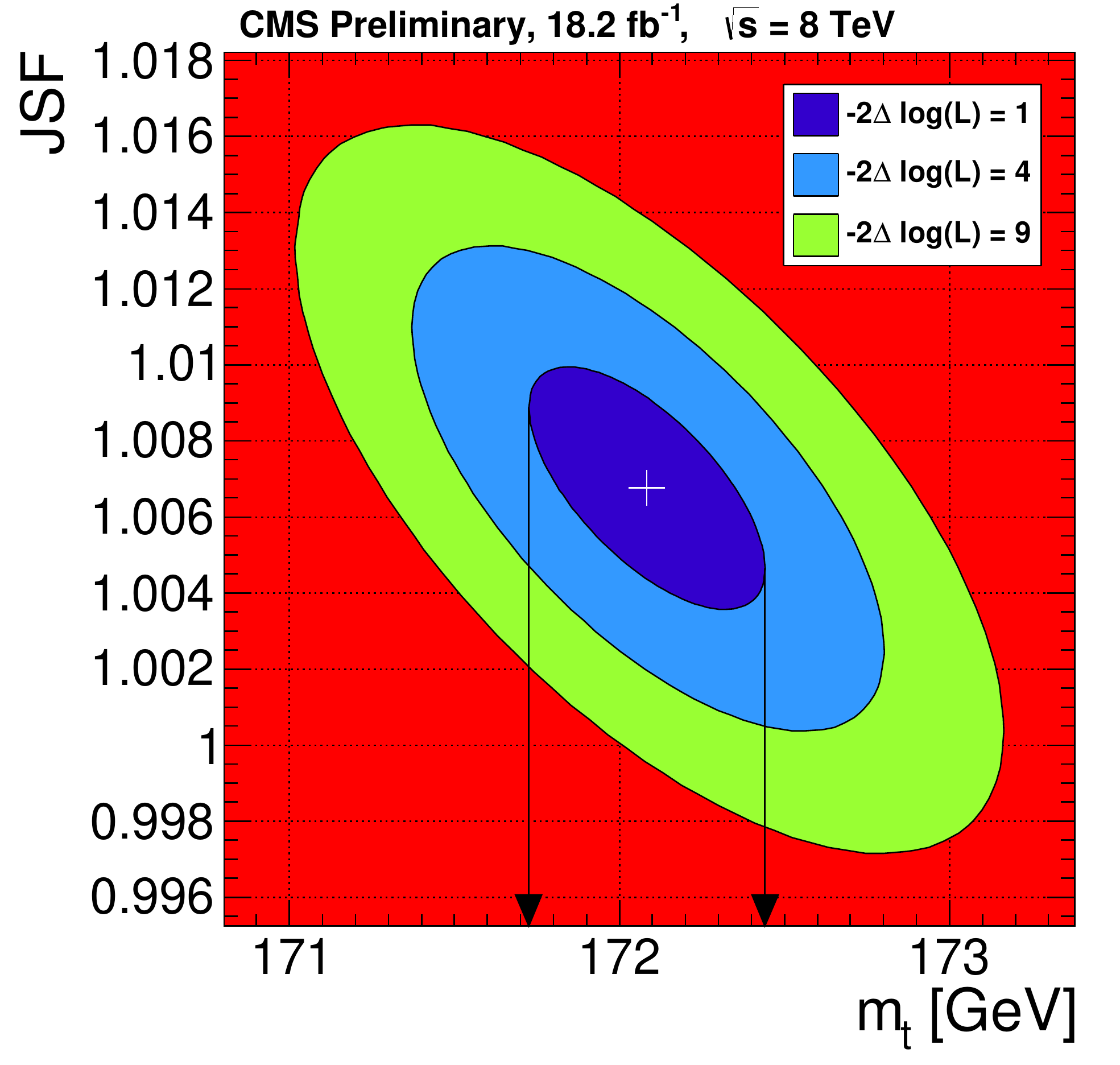}%
  \caption{Likelihood contours for the 2D fit (blue) showing the correlation of the \mt\ fit parameter with the JES in (left) the lepton+jets channel, and (right) the all-hadronic channel at CMS \protect{\cite{cms-ljets,cms-allhad}}.}
  \label{fig:cms_corr}
\end{figure}

Template shapes are obtained for the \mtfit\ and \mwreco\ distributions in three categories of jet-parton permutations -- \textit{correct}, \textit{wrong}, and \textit{unmatched} -- and are used to construct an event-by-event likelihood function with \mt\ and JES as free parameters.  All the event-by-event likelihoods are combined and maximized in order to measure the two parameters of interest (Figure~\ref{fig:cms_ljets_templates}).  The 2D fit in the combined electron and muon channels yields the result:
\begin{align*}
  \mt^{\text{2D}} &= 172.04 \pm 0.19 \text{ (stat+JES)} \pm 0.75 \text{ (syst) GeV}, \\
  \text{JES}^{\text{2D}} &= 1.007 \pm 0.002 \text{ (stat)} \pm 0.012 \text{ (syst)}.
\end{align*}
Comparing to a 1D fit with no in-situ JES calibration, the 2D approach is successful in reducing the systematic uncertainties due to the \pt\ and $\eta$-dependent JES from 1.17 to 0.18 GeV.  The likelihood contour showing the correlation between the \mt\ and JES fit parameters is shown in Figure~\ref{fig:cms_corr}.

In the all-hadronic final state, a dataset is used with an integrated luminosity of 18.2 fb$^{-1}$ and center-of-mass energy of 8 TeV \cite{cms-allhad}.  Events are selected with at least six jets and at least 2 b tags.  The analysis strategy is similar to that in the lepton+jets channel.  It includes a kinematic fit to increase the fraction of correct jet-parton assignments and to improve the \mtreco\ resolution.  The value of \mt\ and JES are extracted in a 2D likelihood fit.  A data driven event mixing technique is implemented to model the shape of the QCD multijet background.  Figure~\ref{fig:cms_kinfit} (right) shows the \mtfit\ distribution observed data and MC with the estimated breakdown of signal and background events.  The 2D fit in the all-hadronic channel yields:
\begin{align*}
  \mt^{\text{2D}} &= 172.08 \pm 0.36 \text{ (stat+JES)} \pm 0.83 \text{ (syst) GeV}, \\
  \text{JES}^{\text{2D}} &= 1.007 \pm 0.003 \text{ (stat)} \pm 0.011 \text{ (syst)},
\end{align*}
where, comparing to the 1D fit, the uncertainties due to the \pt\ and $\eta$-dependent JES are reduced from 0.86 to 0.28 GeV.  The likelihood contour showing the correlation between the \mt\ and JES fit parameters is shown in Figure~\ref{fig:cms_corr}.

\section{ATLAS MEASUREMENTS}
\label{sec:ATLAS}
\begin{figure}
  \includegraphics[width=0.32\textwidth]{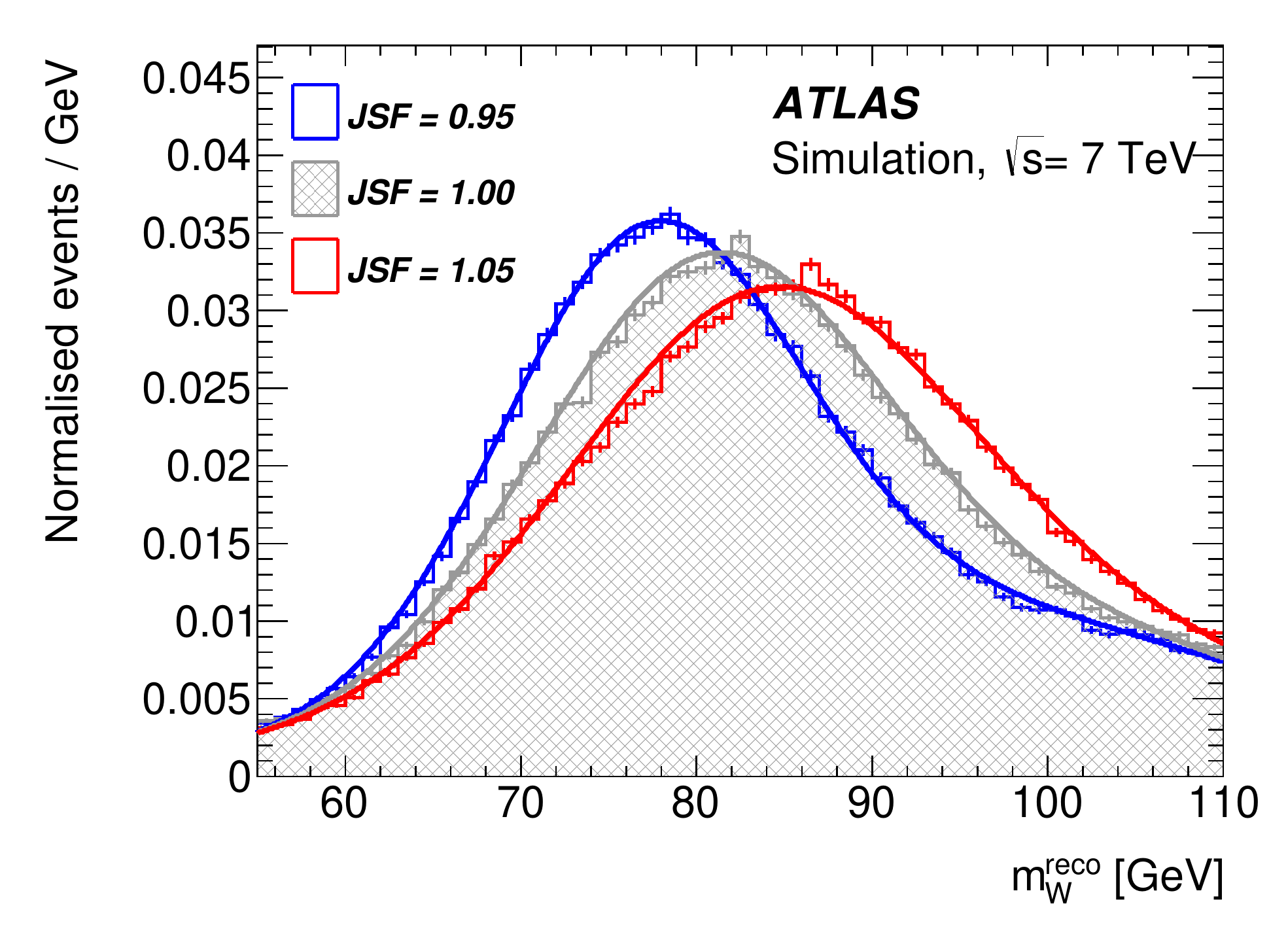}%
  \includegraphics[width=0.32\textwidth]{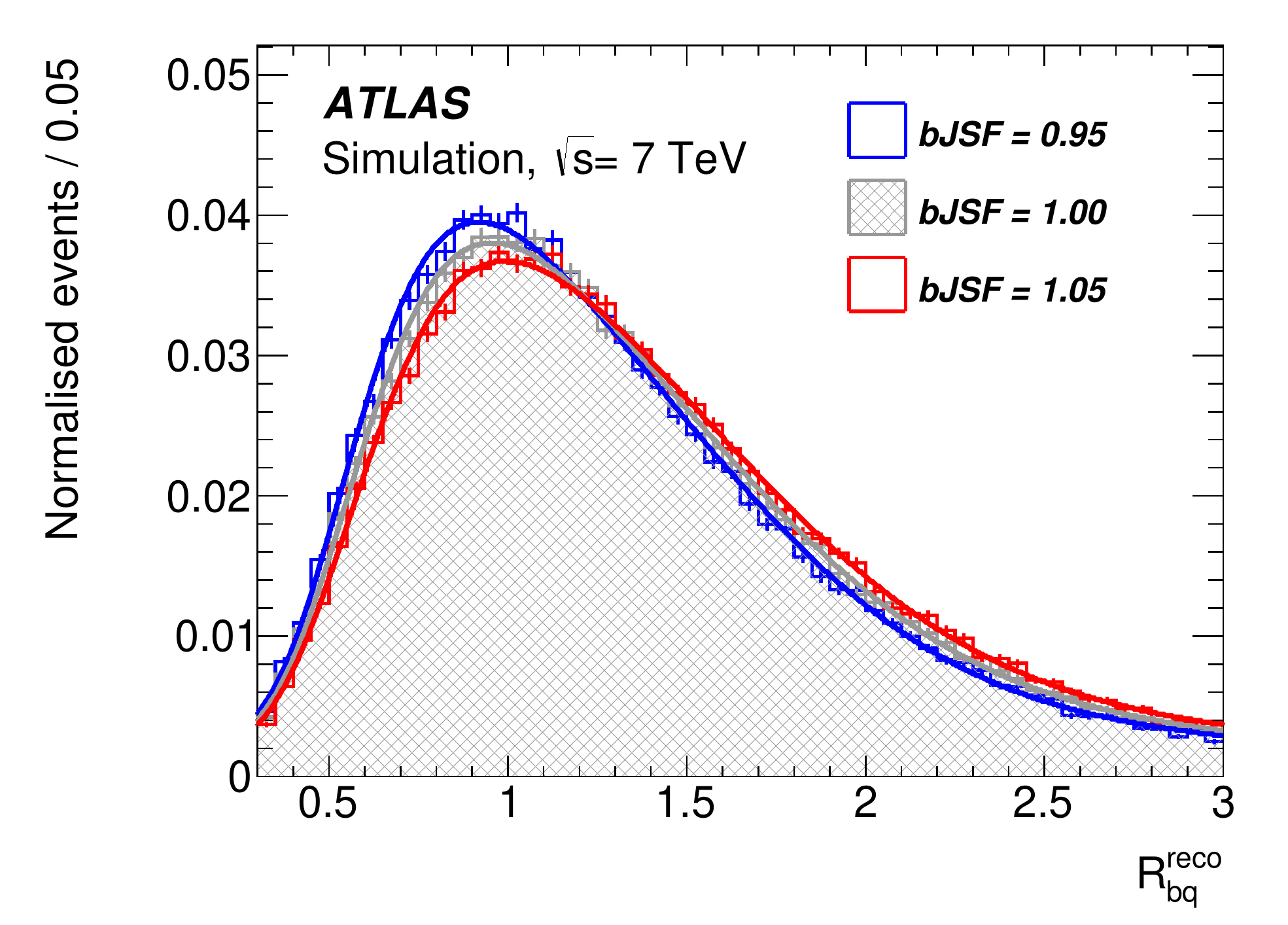}%
  \includegraphics[width=0.32\textwidth]{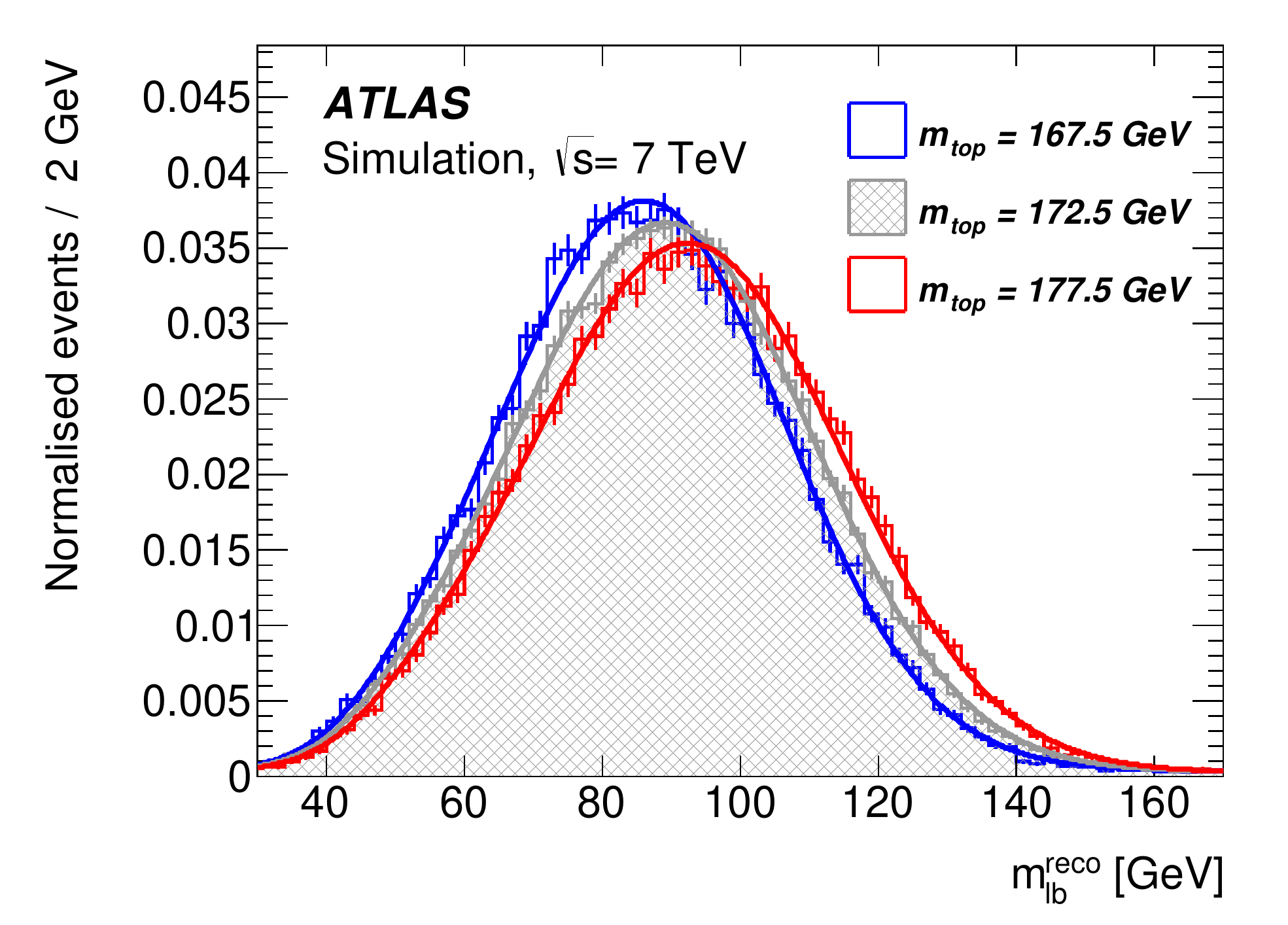}%
  \caption{(Left) \mwreco\ and (center) \rbq\ shapes as a function of the JES and bJES, respectively, in the ATLAS lepton+jets channel.  (Right) \mbl\ shape in the dilepton channel at ATLAS as a function of the top mass \protect{\cite{ATLAS-ttbar}}.}
  \label{fig:atlas_ttbar_templates}
\end{figure}

\begin{figure}
  \includegraphics[width=0.32\textwidth]{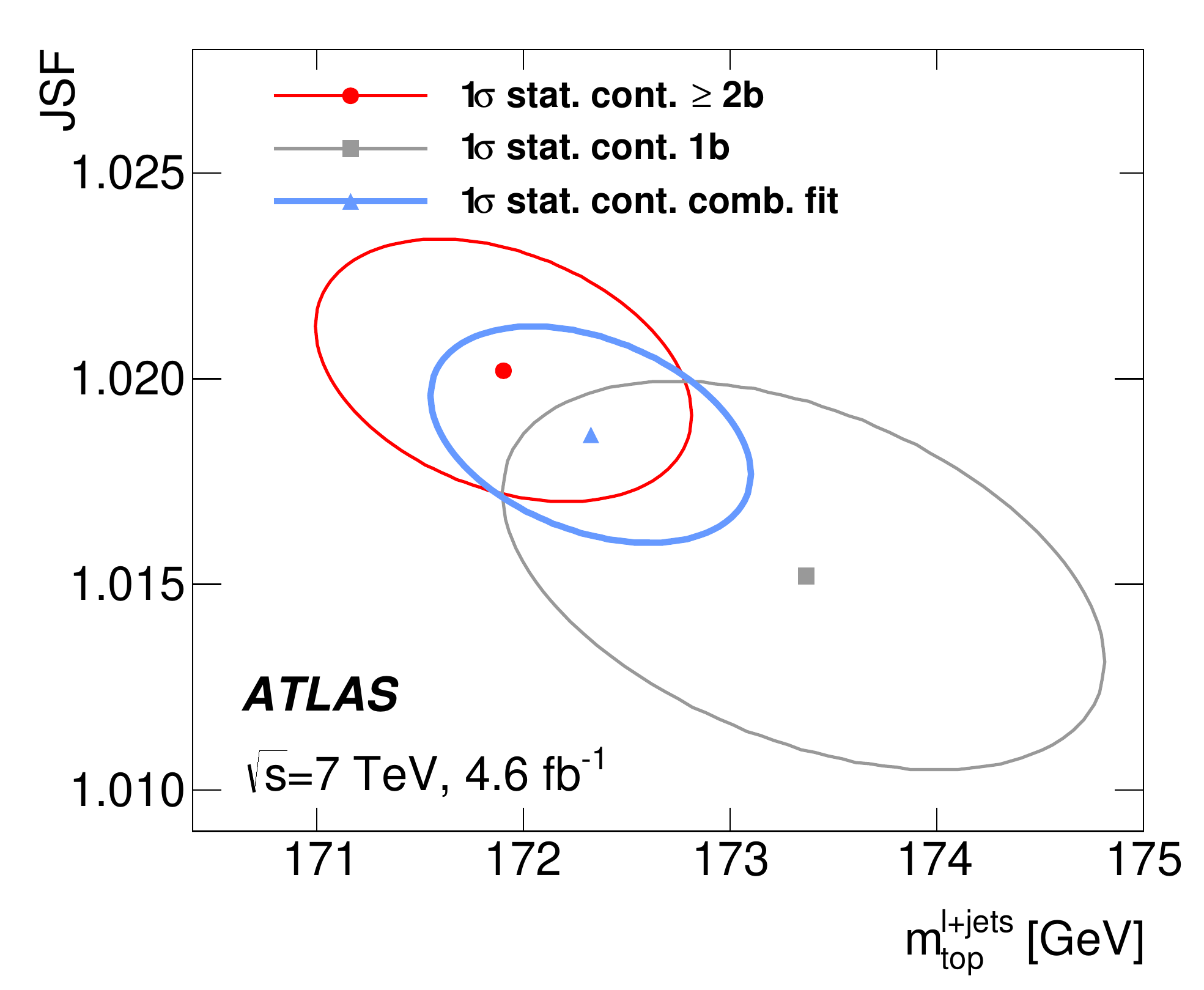}%
  \includegraphics[width=0.32\textwidth]{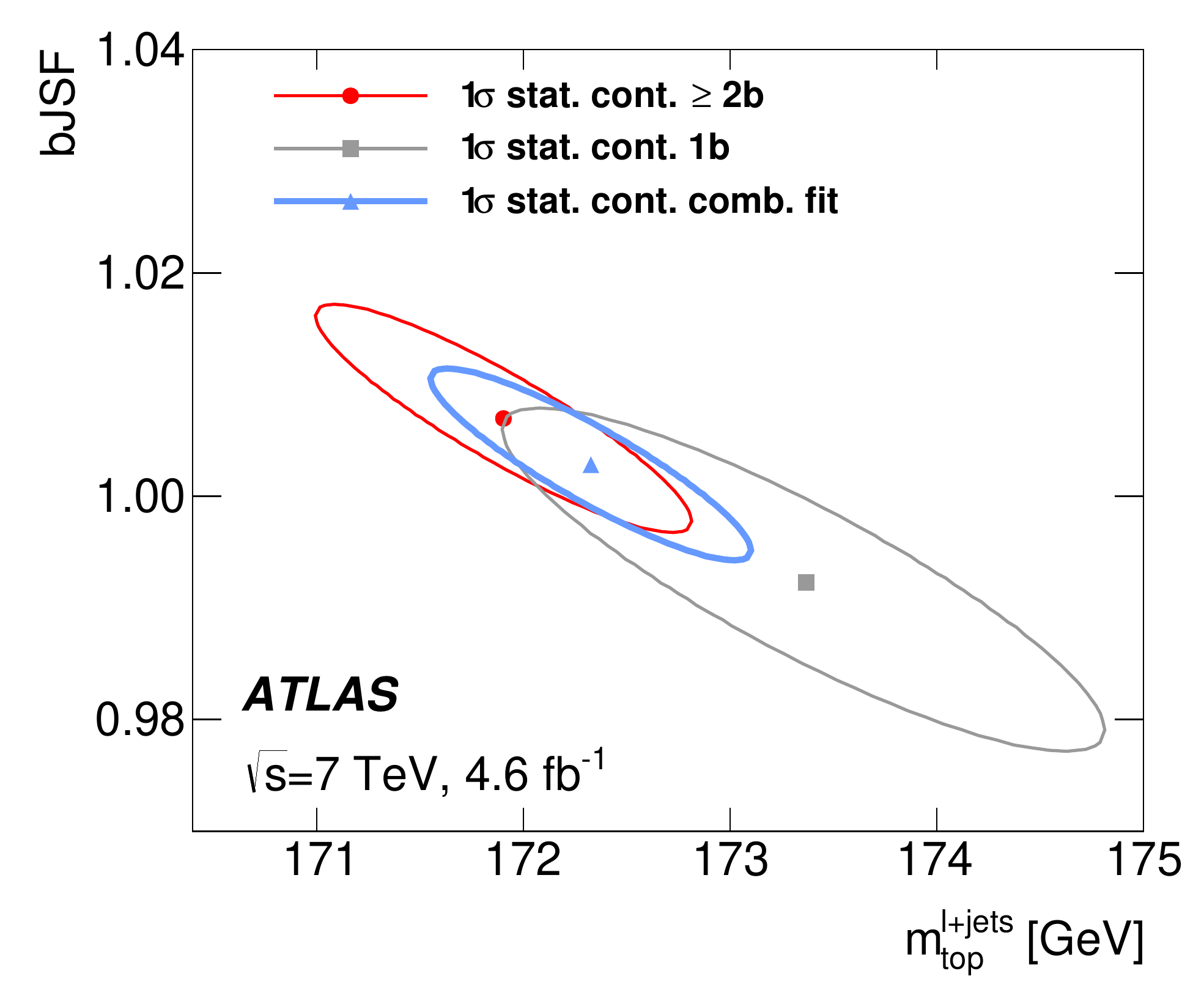}%
  \includegraphics[width=0.32\textwidth]{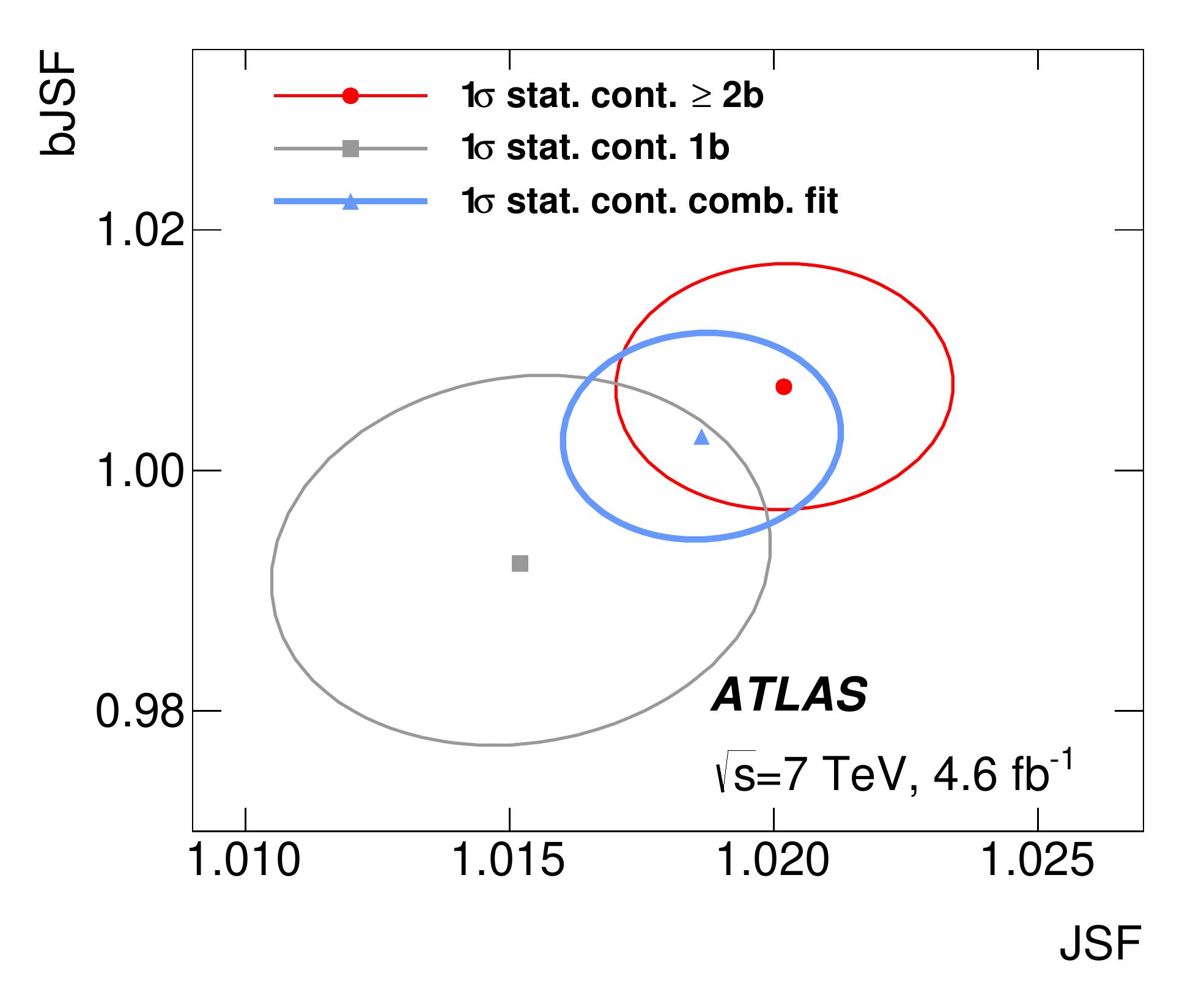}%
  \caption{Likelihood contours showing the correlation of the \mt\ fit parameter with (left) the JES, (center) the bJES, and (right) the correlation between JES and bJES in the lepton+jets channel at ATLAS \protect{\cite{ATLAS-ttbar}}.}
  \label{fig:atlas_corr}
\end{figure}

\begin{figure}
  \includegraphics[width=0.32\textwidth]{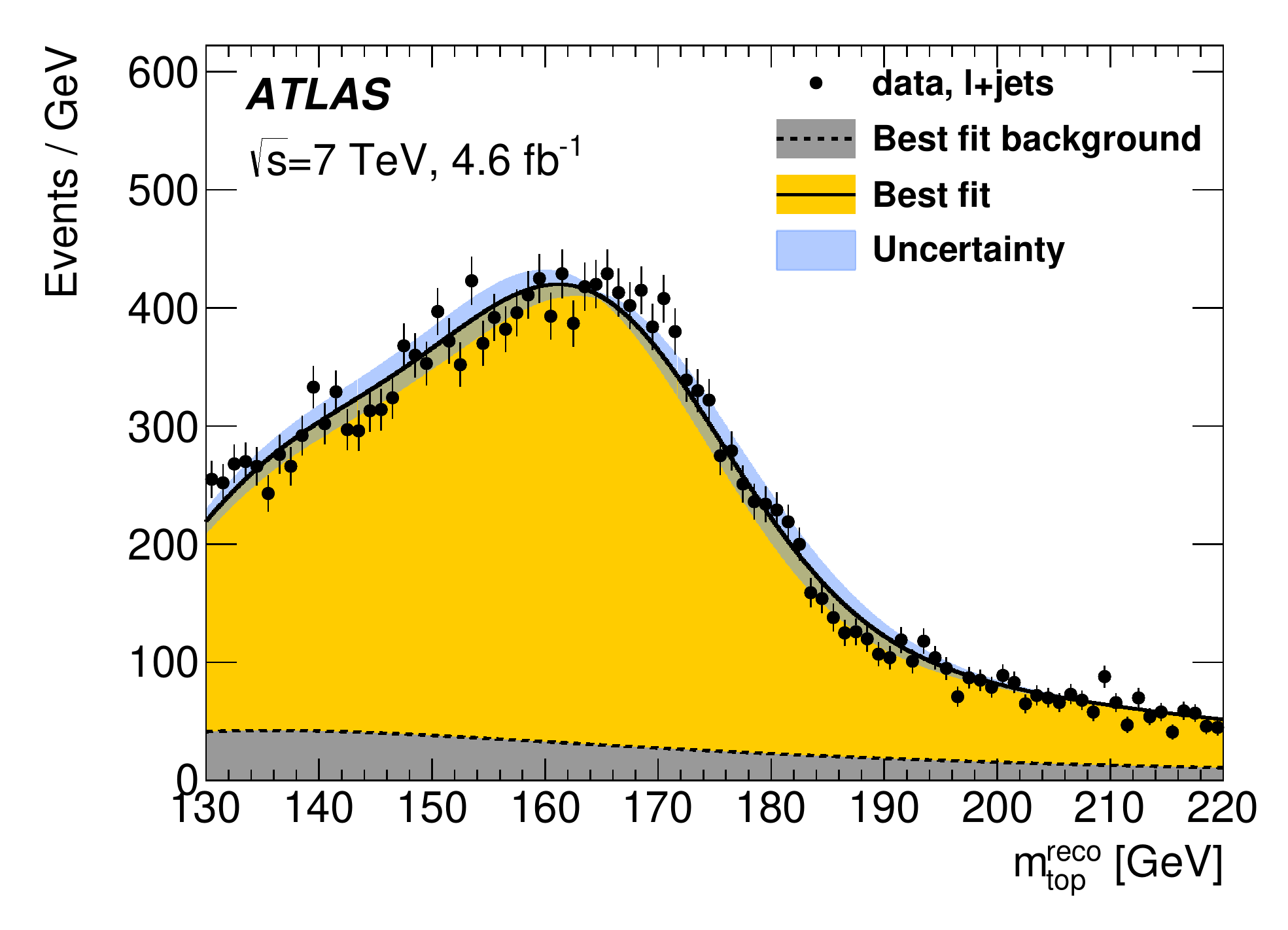}%
  \includegraphics[width=0.32\textwidth]{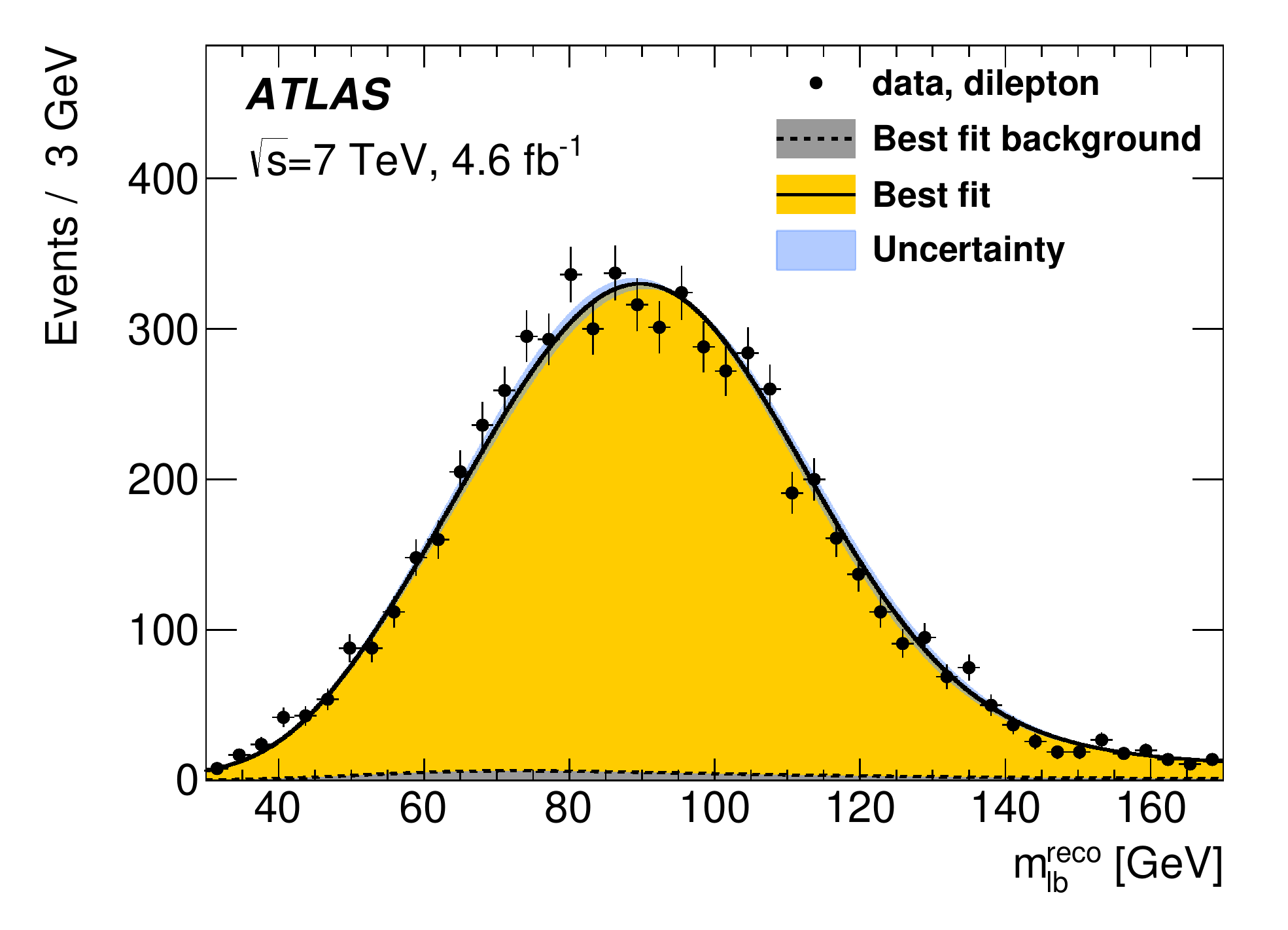}%
  \includegraphics[width=0.32\textwidth]{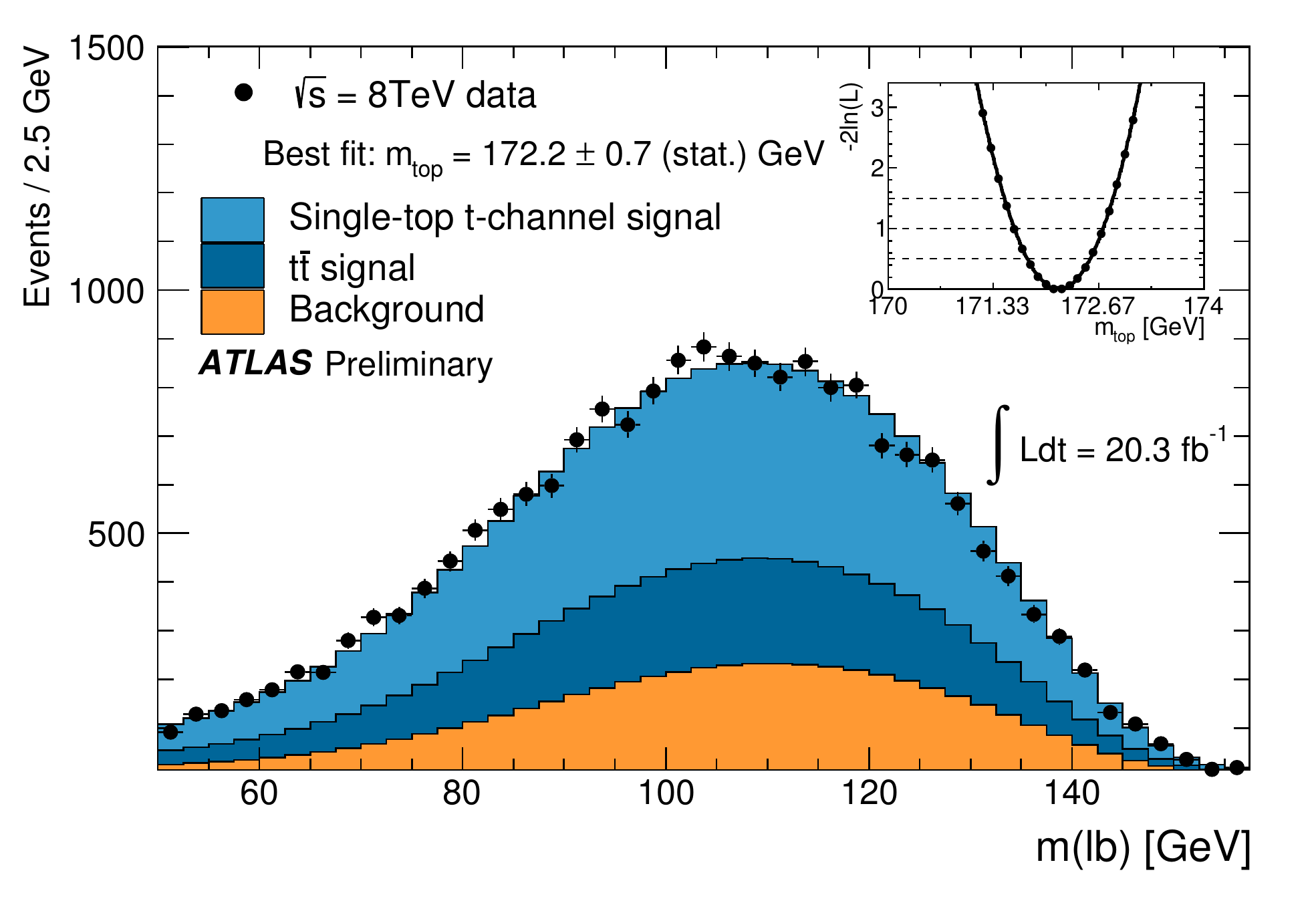}%
  \caption{Best-fit distributions of (left) \mtreco\ in the lepton+jets channel, (center) \mbl\ in the dilepton channel, and (right) \mbl\ in a topology enriched with t-channel single top events at ATLAS \protect{\cite{ATLAS-ttbar,ATLAS-CONF-2014-055}}.}
  \label{fig:atlas_bestfit}
\end{figure}

In the lepton+jets channel, a measurement performed by ATLAS features a 3D likelihood fit, constraining the \pt\ and $\eta$-dependent JES in addition to the flavor-dependent JES component \cite{ATLAS-ttbar}.  The flavor-dependent JES uncertainty accounts for differences between b jets and udsg jets, and is one of the leading systematic uncertainties in top mass measurements where the \pt\ and $\eta$-dependent JES is constrained.
Events are selected from the ATLAS 7 TeV dataset with an integrated luminosity of 4.6 fb$^{-1}$.  Event selection requirements include one electron or muon, at least four jets, at least 1 b tag, as well as requirements on the MET and $\mathrm{m}_\text{T}^{\mathrm{W}}$.  Separate fits are conducted using events with exactly 1 b tag, and 2 or more b tags.  The \ttbar\ system is reconstructed with a kinematic likelihood fit similar to that implemented by CMS.  Templates are constructed to model the shapes of the \mtreco\ and \mwreco\ distributions, as well as the \rbq\ variable, defined by:
\begin{align}
  \label{eq:Rbq}
  \mathrm{R}^{\text{reco,1b}}_{\mathrm{bq}} = \frac{\pt^{\mathrm{b}}}{\frac{1}{2}(\pt^{\mathrm{W}_{\text{jet1}}}+\pt^{\mathrm{W}_{\text{jet2}}})}, \qquad
  \mathrm{R}^{\text{reco,2b}}_{\mathrm{bq}} = \frac{\pt^{\mathrm{b}_{\text{had}}}+\pt^{\mathrm{b}_{\text{lep}}}}{\pt^{\mathrm{W}_{\text{jet1}}}+\pt^{\mathrm{W}_{\text{jet2}}}}.
\end{align}
In the 3D likelihood fit, \mtreco\ and \mwreco\ are sensitive to \mt\ and the overall (\pt\ and $\eta$-dependent) JES, respectively.  Because \rbq\ is constructed as a ratio of b jet to udsg jet momenta, its sensitivity lies mostly in the \textit{relative} JES of b jets compared to udsg jets, denoted the bJES.  This is illustrated in the distribution shapes shown in Figure~\ref{fig:atlas_ttbar_templates}.  The 3D fit yields:
\begin{align*}
  \mt^{\text{3D}} &= 172.33 \pm 0.75 \text{ (stat+JES)} \pm 1.02 \text{ (syst) GeV}, \\
  \text{JES}^{\text{3D}} &= 1.019 \pm 0.003 \text{ (stat)} \pm 0.027 \text{ (syst)}, \\
  \text{bJES}^{\text{3D}} &= 1.003 \pm 0.008 \text{ (stat)} \pm 0.023 \text{ (syst)}.
\end{align*}
Here, the method is still limited by the available statistics in the 7 TeV dataset, and the precision is expected to improve in future iterations of the measurement.  Likelihood contours showing the correlations between the three fit parameters are shown in Figure~\ref{fig:atlas_corr}.  The best-fit distribution of \mtreco\ is shown in Figure~\ref{fig:atlas_bestfit}.

In the dilepton channel, ATLAS performs a measurement using the \mbl\ observable, defined as the invariant mass between b jet, lepton pairs \cite{ATLAS-ttbar}.  The \mbl\ variable provides an alternative to the full reconstruction of the \ttbar\ system in a channel that is kinematically underconstrained.  In this channel, the top mass cannot be reconstructed in a single event, but over an ensemble of events the distribution of \mbl\ is sensitive to \mt\ through its shape and the location of its endpoint (Figure~\ref{fig:atlas_ttbar_templates}).  Events are selected from the ATLAS 7 TeV dataset with an integrated luminosity of 4.6 fb$^{-1}$.  They are required to contain two isolated oppositely-charged electrons or muons, at least 2 jets, and one or two b tags.  The shape of \mbl\ is modeled separately for the \mt-dependent signal and small \mt-independent background.  The measurement yields:
\begin{align*}
  \mt = 173.79 \pm 0.54 \text{ (stat)} \pm 1.30 \text{ (syst) GeV}.
\end{align*}
Here the JES is the largest contributor to the total systematic uncertainty, giving 1.01 GeV (JES + bJES).  The best-fit distribution of \mbl\ is shown in Figure~\ref{fig:atlas_bestfit}.

To complement top mass measurements utilizing \ttbar\ topologies, ATLAS has performed a measurement using a sample enriched in t-channel single top events \cite{ATLAS-CONF-2014-055}.  Events are selected from the ATLAS 8 TeV dataset with an integrated luminosity of 20.3 fb$^{-1}$.  The base event selection includes a single isolated lepton, high MET, at least two jets, and exactly one b tag.  To further improve the purity of the signal sample, a neural network is constructed with 12 input variables.  An appropriate cut on the neural network output delivers a signal purity of 50\%, where the remaining events are mostly due to \ttbar\ and other single top production modes, as well as a W+jets background.  The analysis uses the \mbl\ observable as in the dilepton channel, where shapes for the single top + \ttbar\ signal and the non-top backgrounds are modeled separately (Figure~\ref{fig:atlas_bestfit}).  A likelihood fit to the \mbl\ distribution yields:
\begin{align*}
  \mt = 172.2 \pm 0.7 \text{ (stat)} \pm 2.0 \text{ (syst) GeV}.
\end{align*}
The JES accounts for the largest source of systematic uncertainty, with the combined uncertainty of 1.5 GeV due to JES and bJES.  The best-fit distribution of \mbl\ is shown in Figure~\ref{fig:atlas_bestfit}.

\section{COMBINATIONS}
\label{sec:combination}
\begin{figure}[h]
  \includegraphics[width=0.75\textwidth]{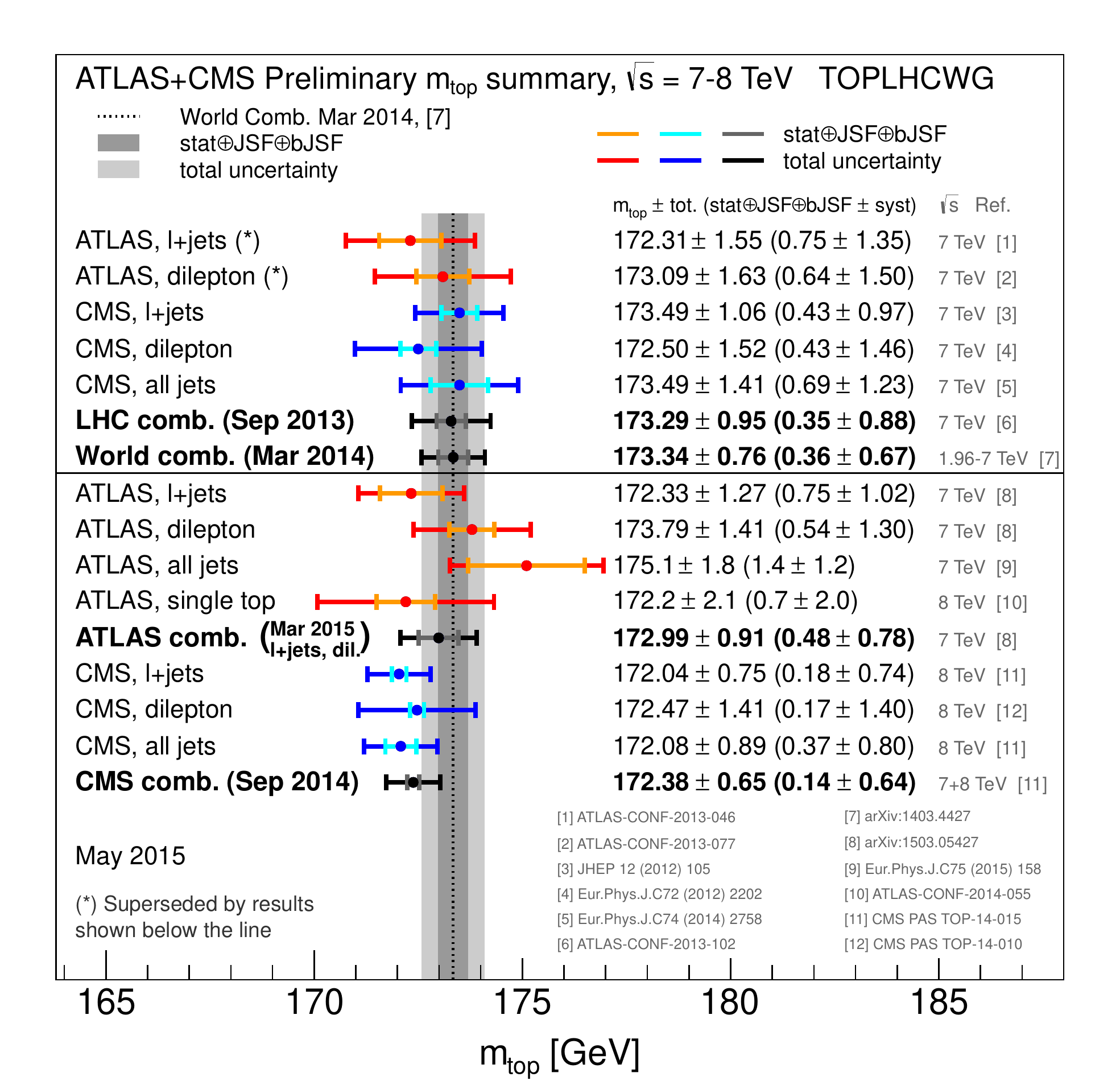}%
  \caption{Recent combinations of top quark mass measurements by the LHC and Tevatron experiments \protect{\cite{toplhcwg}}.}
  \label{fig:comb}
\end{figure}

The results outlined in these proceedings enter into combinations that use the Best Linear Unbiased Estimator (BLUE) method, which accounts for correlations between all sources of measurement uncertainty \cite{BLUE}.  An overview of recent combinations is given in Figure~\ref{fig:comb}.  Recent single-experiment combinations by ATLAS and CMS achieve a sub-GeV precision on \mt\ \cite{cms-comb,ATLAS-ttbar}.  At the time of the LHCP 2015 conference, the most precise combination at the LHC was conducted using measurements in the lepton+jets, all-hadronic, and dilepton channels at CMS, yielding a result of $\mt = 172.38 \pm 0.66$ GeV \cite{cms-comb}.

\section{SUMMARY}
\label{sec:conclusion}
After a productive Run 1 of the LHC, the ATLAS and CMS experiments have delivered measurements of the top quark mass in three \ttbar\ channels, and topologies enriched in t-channel single top.  In the lepton+jets and all-hadronic channels, both measurements use techniques that reduce their sensitivity to the JES systematic uncertainties.  The CMS measurements implement a 2D fit using the \mtfit\ and \mwreco\ observables.  The ATLAS lepton+jets measurement implements a 3D fit, adding the observable \rbq\ in order to also constrain the flavor-dependent JES uncertainty.  In the dilepton and t-channel single top channels, ATLAS uses the \mbl\ observable for event reconstruction.  Recent combinations by both CMS and ATLAS achieve sub-GeV precision on the value of the top quark mass.


\bibliographystyle{hunsrt}
\bibliography{proceedings}

\end{document}